\documentclass[reprint,superscriptaddress,amsmath,amssymb,aps,pre,nofootinbib]{revtex4-2}
\usepackage[bookmarks, colorlinks, breaklinks, pdfusetitle]{hyperref}  
\hypersetup{linkcolor=blue, citecolor=blue, filecolor=black, urlcolor=blue}

\usepackage[T1]{fontenc}
\usepackage{lmodern}
\usepackage[acronym]{glossaries}
\usepackage{newtxtext}
\usepackage[caption=false]{subfig}
\usepackage{leftidx}

\usepackage{tabularx}
\newcolumntype{Y}{>{\centering\arraybackslash}X}
\newcolumntype{s}{>{\hsize=.4\hsize}X}
\newcolumntype{Z}{>{\hsize=.6\hsize}Y}

\usepackage{graphicx}

\usepackage{siunitx}
\usepackage{cleveref}
\crefname{equation}{Eq.}{Eqs.}
\crefname{figure}{Fig.}{Figs.}
\crefname{table}{Table}{Tables}
\crefname{section}{Sec.}{Secs.}

\newcommand{\UVa}{Departamento de F\'{i}sica Te\'{o}rica At\'{o}mica y \'{O}ptica, Universidad de Valladolid, 47011 Valladolid, Spain}
\newcommand{\ULPGC}{iUNAT-Departamento de F\'{i}sica, Universidad de Las Palmas de Gran Canaria, 35017 Las Palmas de Gran Canaria, Spain}
\newcommand{\WIS}{Faculty of Physics, Weizmann Institute of Science, Rehovot 7610001, Israel}

\renewcommand{\l}{\ell}

\newcommand{\ccm}{cm^{-3}}

\usepackage{xspace}
\newcommand{\HeB}{He-$\beta$\xspace}
\newcommand{\ArXVII}{Ar\hspace{2bp}{\sc xvii}\xspace}
\newcommand{\KrXXXV}{Kr\hspace{2bp}{\sc xxxv}\xspace}

\glsdisablehyper
\newacronym{CSM}{CSM}{computer simulation model}
\newacronym{FWHA}{FWHA}{full width at half area}
\newacronym{ICF}{ICF}{inertial confinement fusion}
\newacronym{HED}{HED}{high energy density}
\newacronym{NIF}{NIF}{National Ignition Facility}
\newacronym{FAC}{FAC}{Flexible Atomic Code}
\newacronym{SEIT}{SEIT}{simulation-equivalent interference terms}

\begin{document}

\title{Computer simulations of the Stark effect in the helium-\texorpdfstring{$\beta$}{beta} complex of krypton in ICF conditions}

\author{G. P\'{e}rez-Callejo}
    \email{gabriel.perez.callejo@uva.es}
    \affiliation{\UVa}
\author{E. Stambulchik}
    \email{evgeny.stambulchik@weizmann.ac.il}
    \affiliation{\WIS}
\author{R. Florido}
    \affiliation{\ULPGC}
\author{M. A. Gigosos}
    \affiliation{\UVa}
    
\date{\today}

\begin{abstract}
There is an ongoing interest in using spectroscopy in inertial confinement fusion (ICF) experiments, where dopants such as krypton can provide vital information about the temperature and density of the imploding plasma.
While the most advanced tools for calculating Stark profiles are computer simulation models (CSMs), their application to complex lineshapes under the extreme conditions of ICF experiments is computationally challenging.
In this manuscript, we present results of several CSM realizations applied to the Stark shape of the krypton \HeB line and its satellites at ICF-relevant conditions ($n_e = \qtyrange{1e24}{1e25}{\ccm}$, $T_e=\qty{3}{keV}$).
We demonstrate that codes with the same underlying physics but different numerical approaches yield identical results and analyze the differences in the line profile caused by various physical effects.
\end{abstract}

\maketitle

\section{Introduction}
$K$-shell transitions of He-like species have been widely used to diagnose \gls{HED} plasmas, including those in \gls{ICF} experiments \cite{florido:2011, regan:2013a, hill:2016a, chen:2017a, barrios:2018, slutz:2018a, perezcallejo:2019, perezcallejo:2020, kraus:2021, gao:2022a, hill:2022a, baillygrandvaux:2024}. In particular, \HeB (a transition between the states with the principal quantum number $n = 3$ and $n = 1$ in a two-electron ion) is useful for density measurements.
In many studies of plasma having a temperature of the order of \qty{1}{keV}, \ArXVII \HeB has been used for this purpose \cite{haynes:1996a, woolsey:1997, golovkin:2000, junkel:2000a, golovkin:2002}. However, argon is strongly ionized in hotter plasmas, and higher-$Z$ tracer species must be used for diagnostics.
Indeed, Stark broadening of krypton \HeB was used for inferring the electron density of compressed capsules at the \gls{NIF} \cite{gao:2022a, hill:2022a} and suggested for diagnostics in Laser MegaJoule experiments~\cite{perez-callejo:2022a}.

The \HeB transition is always accompanied by its Li-like satellites with one spectator electron occupying various orbitals.
These satellites partially overlap with the \HeB line shape, requiring one to model and analyze the entire \HeB complex for reliable diagnostics. 

The satellite transitions significantly increase the complexity of the lineshape calculations. For example, Li-like satellites $1s n^\prime \l^\prime 3\l \rightarrow 1s^2 n^\prime \l^\prime$ with $n^\prime = 2, 3,$ and $4$ ($0 \le \l \le 2$, $0 \le \l^\prime < n^\prime$) comprise 69, 68 and 209 atomic levels with 841, 760, and 6206 non-zero electric dipole matrix elements, respectively---not counting for the projection of the total angular momentum.
The increased complexity makes applying Stark-broadening models~\cite{gigosos:2014a} computationally challenging in general and becomes nearly prohibitive for computer simulations~\cite{stambulchik:2010a}. In fact, while computer simulations for the Stark broadening of the Kr \HeB line itself were recently produced \cite{hill:2022a}, no computer simulation results for the Li-like satellite emission have been published---neither for this, nor for any other atomic system.

Recently, \citet{gallardo-diaz:2024a} performed Stark broadening calculations of Kr \HeB and its $n = 2,3$ Li-like satellites using the so-called ``standard theory'' approximation~\cite{griem:1974a} that assumes static ions and impact electrons.
In particular, the effect of the interference terms in the electron impact operator of satellite transitions~\cite{iglesias:2010a} was analyzed and found to be minor, in agreement with a similar study of the Ar \HeB complex~\cite{mancini:2013a}.
When justified, omitting the interference terms in ``standard theory'' models allows one to reduce the electron-operator matrix dimensions several times, significantly improving the computational speed.

In the present study, we report on computer simulations applied, for the first time, to the same atomic system (Kr) under conditions relevant to \gls{ICF}: tracer amounts of Kr in a deuterium plasma, with electron densities in the range from \qtyrange{1e24}{1e25}{\ccm} and temperature of \qty{3}{keV}. We model the \HeB line and its $n=2$ and $n=3$ Li-like satellites using different codes with different physical, mathematical, and numerical approaches. We show the effect of four different levels of particle interaction in the model: independent particles, interaction with the emitter, full $N$-body interaction, and $N$-body interaction with recombination broadening. Additionally, we present a hybrid code combining elements of computer simulations and standard theory. We also present the \textit{simulation-equivalent interference terms} (SEIT), an approach based on the concept of \textit{interference terms} from impact theory of Stark broadening, in the framework of computer simulations. We observe that the SEIT have a minor effect on the shape of the $n=2$ satellites. However, it is not negligible in the case of the $n=3$ satellite transitions, contrary to the recent findings~\cite{gallardo-diaz:2024a}. We discuss the origin of this discrepancy.

This manuscript is structured as follows. In \cref{sec:Codes}, we describe the physical framework for the codes used in this analysis. The results obtained for the \HeB line of Kr and its $n=2$ and $n=3$ satellites are presented and discussed in \cref{sec:Results}.
\cref{sec:IT} is devoted to the analysis of the interference terms.
Finally, \cref{sec:Conclusion} summarizes the primary outcomes of this work and discusses future paths.

\begin{table*}
\caption{\label{table:CodeComparison} Summary of the characteristics of the different codes presented in this work}
\begin{tabularx}{0.9\textwidth}{s|Y Y Z Y}
\textbf{Code} & \textbf{Particle interaction} & \textbf{Correlation function} & \textbf{Ion field} & \textbf{Electron field} \\ \hline
SIMULA & None (straight paths) & \cref{eqn:TimeDomain} & From simulation & From simulation \\
SIMULAm & None (straight paths) & \cref{eqn:TimeDomain} & From simulation & Following standard theory \\
SimU & With Debye-shielded emitter & \cref{eqn:FreqDomain}  & From simulation & From simulation \\
SimU$_\text{SP}$ & None (straight paths) & \cref{eqn:FreqDomain} & From simulation & From simulation \\
DinMol & Full Coulomb dynamics & \cref{eqn:TimeDomain} & From simulation & From simulation
\end{tabularx}
\end{table*}

\section{Code description}
\label{sec:Codes}

This paper presents results obtained with three different codes---SIMULA~\cite{gigosos:2014b}, SIMULAm, and SimU~\cite{stambulchik:2006a}.
All semiclassical \glspl{CSM}, starting from the pioneering work of \citet{stamm:1979a}, share the same approach:
the motion of charged plasma particles is modeled by numerically solving a classical $N$-body problem, while their fields affect the evolution of the atomic system of the emitter, treated quantum-mechanically.

While the idea behind the codes is the same, they operate in different ways. In SIMULA and SIMULAm, the particles are considered independent of each other and, therefore, move following rectilinear trajectories. On the other hand, SimU considers the perturbers to be independent, but it explicitly solves their interaction with the emitter, and, therefore, particles move following hyperbolic trajectories if a Coulomb potential is assumed. For the simulations presented in this work, Debye screening at the appropriate plasma conditions was accounted for, which results in more general curves for the particle trajectories. Evidently, if the emitter were neutral, all codes would reproduce the same trajectories; however, that is not the case for the \HeB line of Kr, where the emitter has a charge $Z=+34$. For direct comparison with SIMULA and SIMULAm, we additionally show results obtained with a modified version of SimU, SimU$_\text{SP}$, which enforces straight-path trajectories by disabling the radiator--perturber interaction.

For one particular case, we also show the results obtained with the full-molecular-dynamics code DinMol \cite{gigosos:2018}. In DinMol, all electrons and ions in the simulation interact with one another, and the complex, multibody Coulomb field determines their trajectories. Owing to this detailed treatment of all particles, DinMol can detect when an electron becomes \textit{bound} to an ion \cite{gonzalezherrero:2025}. Therefore, it can include the so-called recombination width \cite{gigosos:2021}.

An additional difference between the codes is how they calculate the lineshape. SimU and SimU$_\text{SP}$ use the following expression to obtain the spectrum
\begin{equation}
    I(\omega) \propto \frac{1}{T} \sum_{u,l}p_u\left\{ \left| \int_{0}^{T} \langle l | D(t) | u\rangle e^{i\omega t} dt \right|^2  \right\},
    \label{eqn:FreqDomain}
\end{equation}
where $T$ indicates a time interval, $D$ is the atomic dipole moment, $u$ and $l$ indicate the upper and lower states of the transition, respectively, $p_u$ is the probability of the atom to be in the upper state $u$ (i.e., $\langle u |\rho |u\rangle$, with $\rho$ being the density matrix) and $\{\cdot\}$ indicates the average over the ensemble of emitters (realized as an average over several simulation runs). On the other hand, SIMULA, SIMULAm, and DinMol calculate the spectrum as
\begin{equation}
    I(\omega) \propto \text{Re} \int_0^T \left\{\text{Tr}\left[\rho D(t)D(0)\right]\right\} e^{-i\omega t} dt.
    \label{eqn:TimeDomain}
\end{equation}

\begin{figure*}
\centering
\subfloat[\label{fig:HeB_1e24}]{
    \includegraphics[height=0.3\linewidth]{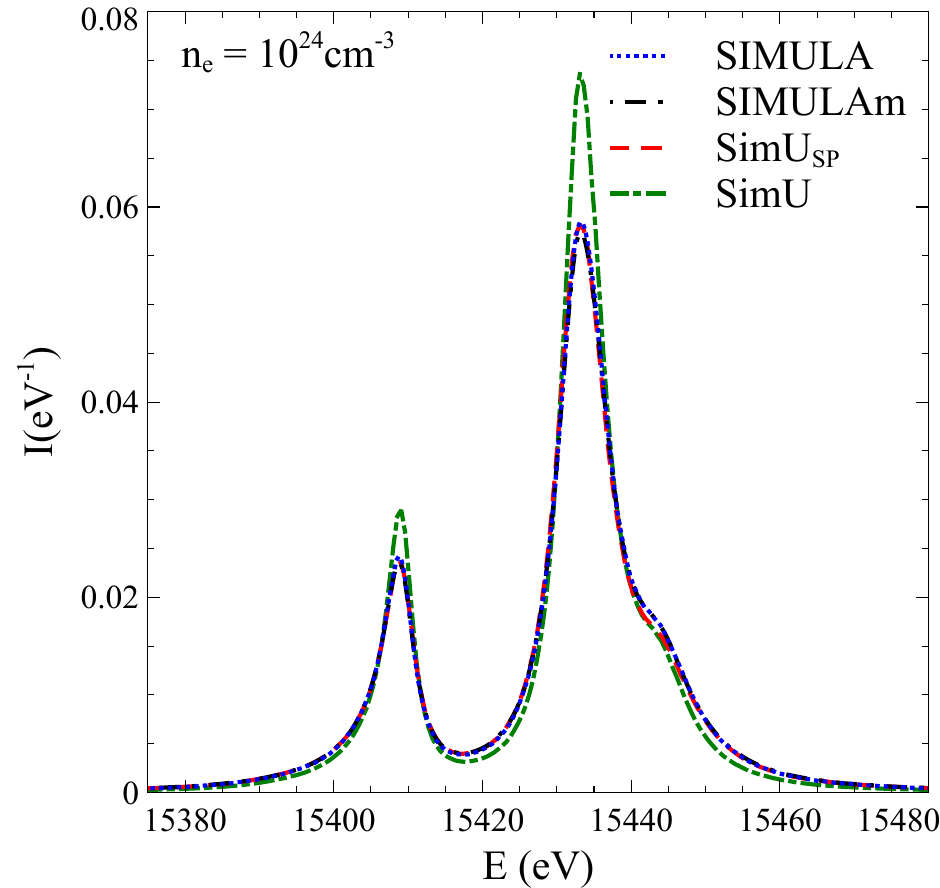}
} 
\subfloat[\label{fig:HeB_3e24}]{
    \includegraphics[height=0.3\linewidth]{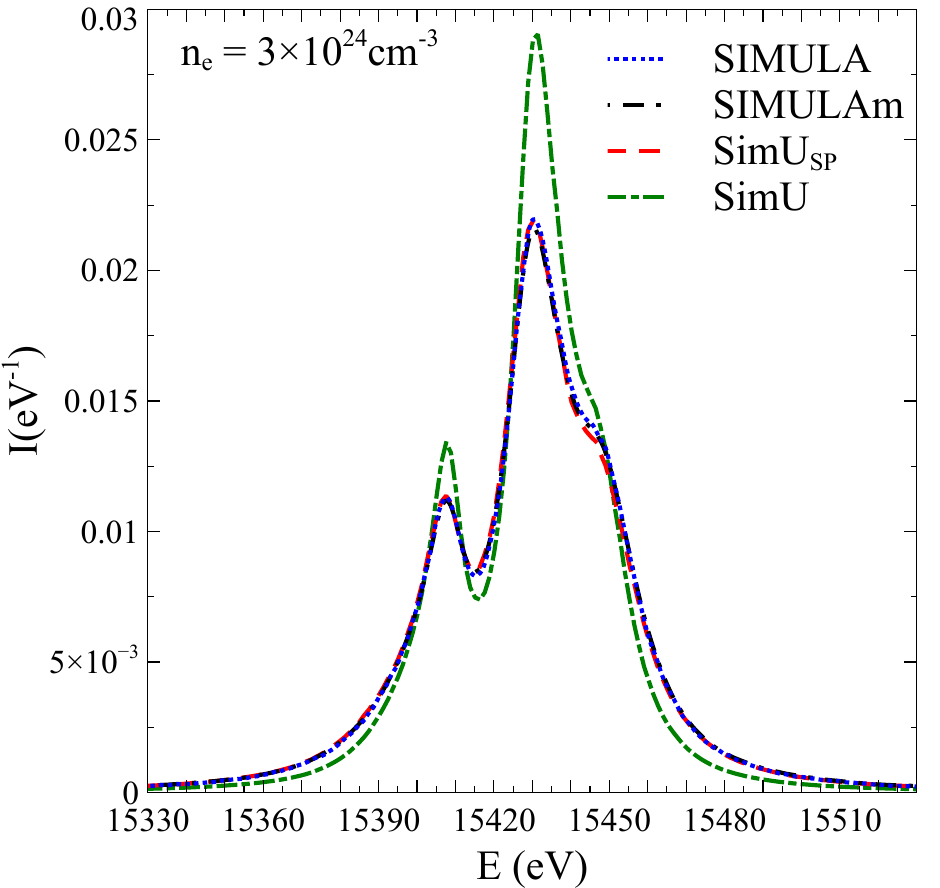}
}
\subfloat[\label{fig:HeB_1e25}]{
    \includegraphics[height=0.3\linewidth]{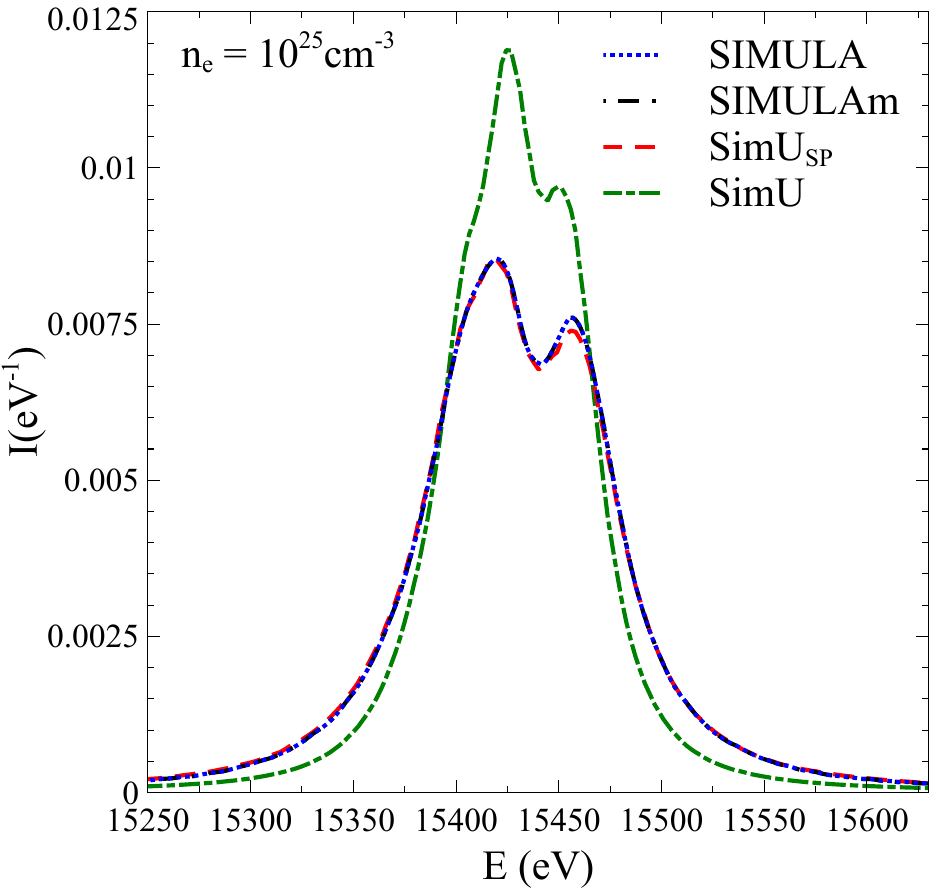}
} 
\caption{A comparison of \KrXXXV \HeB Stark line shapes, calculated for $n_e = \qty{1e24}{\ccm}$ (a), \qty{3e24}{\ccm} (b), and \qty{1e25}{\ccm} (c), using different \glspl{CSM}. $T = \qty{3}{keV}$ is assumed in all cases. The line shapes are area-normalized to unity.}
\label{fig:He}
\end{figure*}

\begin{figure}
\centering
\includegraphics[width=\columnwidth]{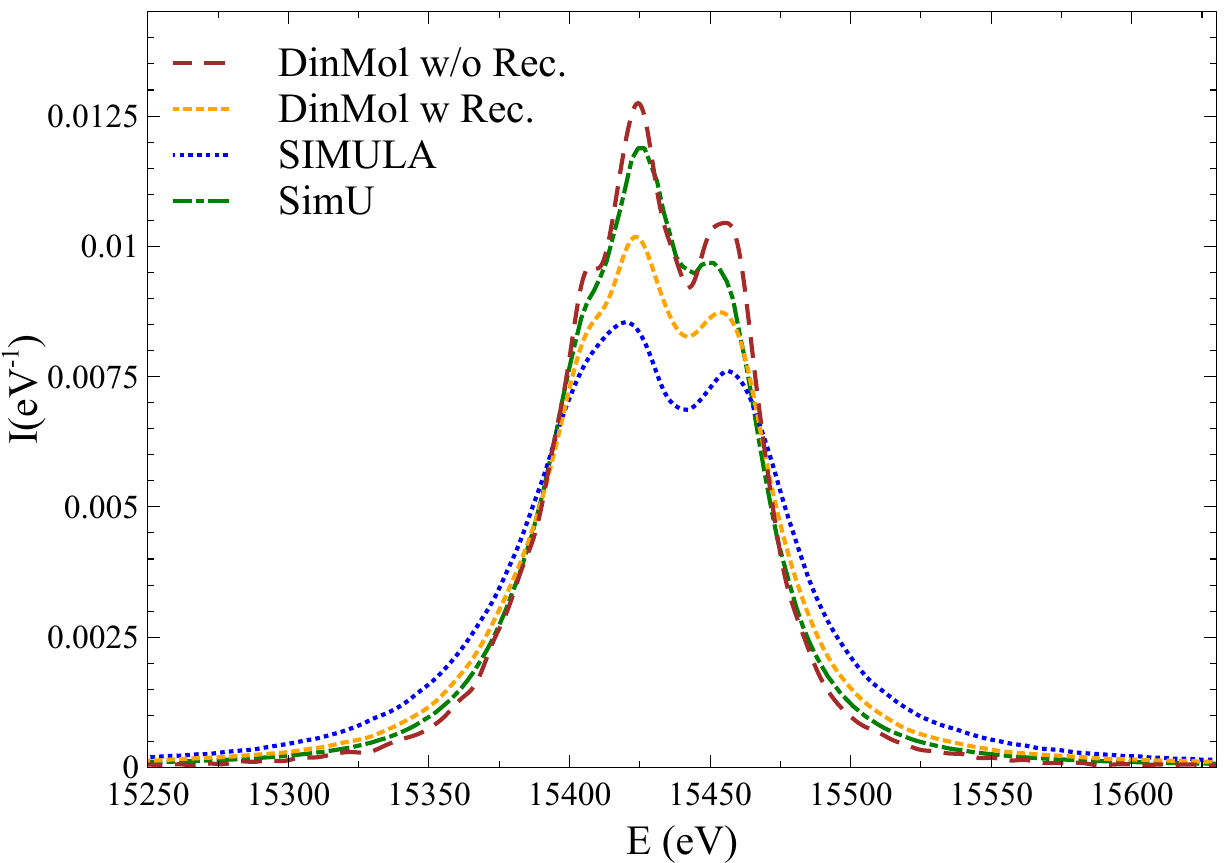}
\caption{\HeB lineshape at $n_e=\qty{1e25}{\ccm}$ and $T=\qty{3}{keV}$, comparing the results SIMULA (independent particles), SimU (interaction with the emitter), and DinMol (full molecular dynamics) with and without recombination effects. Effectively, this illustrates the differences in the lineshape resulting from varying levels of complexity in the simulation.}
\label{fig:HeB_DinMol}
\end{figure}

In the limit where $T\cdot N\rightarrow\infty$, with $N$ being the number of emitters included in the ensemble average, these two expressions are mathematically equivalent (as long as the ergodic theorem holds and the density matrix is diagonal). In practice, however, this limit cannot be attained. Furthermore, the ensemble average must be sufficiently thorough, and certain differences might arise from the statistical and numerical treatment of both equations \cite{stambulchik:2007, rosato:2020}.
Lastly, while SIMULA solves for the evolution of the dipole moment subjected to the effects of perturbing ions and electrons, SIMULAm includes a mathematical simplification that significantly reduces the computational cost of the calculations. In SIMULAm, the ions evolve as usual in the simulation, and the evolution operator corresponding to the ionic field, $U_I(\tau, t)$, is obtained at each timestep (of size $\tau$) of the simulation. However, the effect of the electronic field is included using the \textit{electron broadening operator} from standard theory \cite{griem:1974a}, using the formalism described by \citet{gigosos:2014a}. Effectively, this provides an additional \textit{electronic evolution operator} $U_e(\tau)$---which only depends on the size of the timestep---such that the total evolution operator is $U(t+\tau, t) = U_e(\tau) U_I(\tau, t)$. However, given that the operators $U_e$ and $U_I$ do not commute, the electronic operator must be included at each timestep so that  the evolution operator is written as
\begin{multline}
    U(t+\tau, t) = U_e(\tau)~U_I(\tau, t)~U(t) = \\ 
    U_e(\tau)~U_I(\tau, t)~U_e(\tau)~U_I(\tau, t-\tau)~U(t-\tau)...
\end{multline}
This allows for a much shorter computation time since the time scale of the simulation is that of the ion motion.

The differences between the codes are summarized in \cref{table:CodeComparison}. The atomic data in all codes used in this work were obtained by cFAC~\cite{cfac}, based on the Flexible Atomic Code~\cite{gu:2008}.

Regarding the limitations of this work, the main simplification made here is the widely used dipole approximation, since the full-Coulomb treatment~\cite{gomez:2021, stambulchik:2022a} could not be applied to keep the calculations computationally feasible.
The dipole approximation cannot reproduce the so-called plasma polarization shift~\cite{junkel:2000a}. The other effect of the full-Coulomb treatment, a certain reduction of the electron broadening due to the penetrating collisions, is minor for ``beta'' transitions for which the ion broadening constitutes the major contribution, as will be shown in \cref{sec:IT}.
In addition, all codes used in this work employ a semiclassical treatment for the particles within the plasma. Specifically, while the radiator's evolution is obtained by treating it quantum-mechanically, the particle trajectories are always strictly classical. This imposes certain limitations on the accuracy with which the physical reality is reproduced in the simulation. Examples of effects not included in this treatment are the effects of the emitted radiation on the thermodynamic equilibrium of the plasma (negligible at these high densities and temperatures) or the exchange interaction effects in electron broadening \cite{gomez:2024a}. However, these are expected to be negligible at the conditions of interest (no magnetic fields, $T>\qty{1}{keV}$).

\section{Results}
\label{sec:Results}

\subsection{He-\texorpdfstring{$\beta$}{beta} line}

\begin{figure}
\centering
\includegraphics[width=\columnwidth]{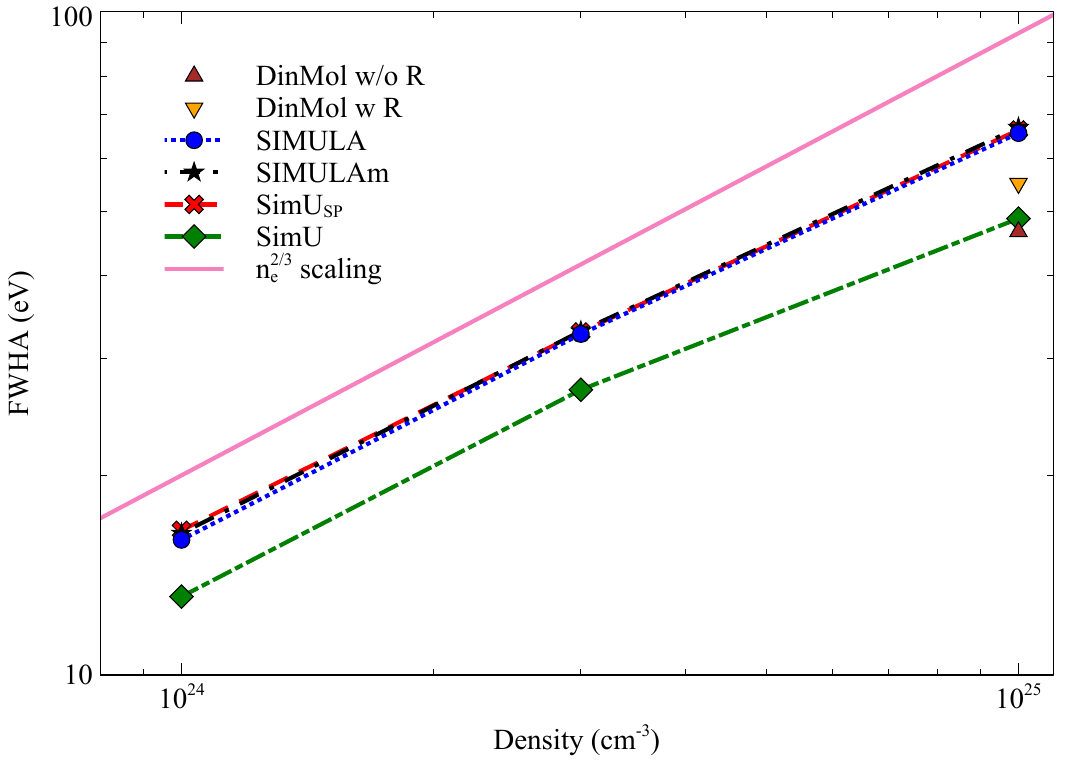}
\caption{Comparison of the \gls{FWHA} obtained at the different density conditions by the \glspl{CSM} presented here for the \HeB line. A solid-pink line indicating the typical $n_e^{2/3}$ scaling has been added to guide the eye. The plot axes are in logarithmic scale, so the $n_e^{2/3}$ scaling is a straight line.}
\label{fig:FWHA}
\end{figure}

A comparison of the area-normalized \HeB line shapes calculated by different models at $T=\qty{3}{keV}$ and electron densities ranging from \qtyrange{1e24}{1e25}{\ccm} for tracer amounts of krypton in a deuterium plasma is shown in \cref{fig:He}. The spectral range has been adjusted in each subfigure. Note that since the profiles are area-normalized, a narrower lineshape results in a higher peak, highlighting the differences between the codes.

\begin{figure*}
\centering
\subfloat[\label{fig:HeBe2_1e24}]{
    \includegraphics[height=0.3\linewidth]{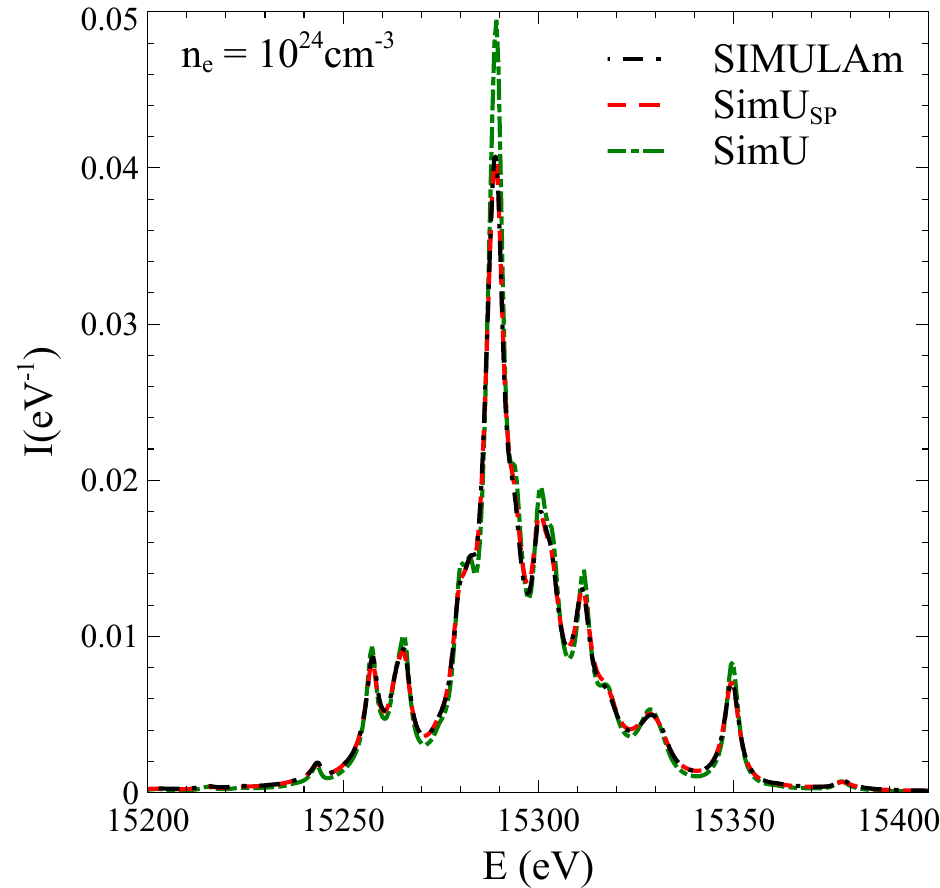}
} 
\subfloat[\label{fig:HeBe2_3e24}]{
    \includegraphics[height=0.3\linewidth]{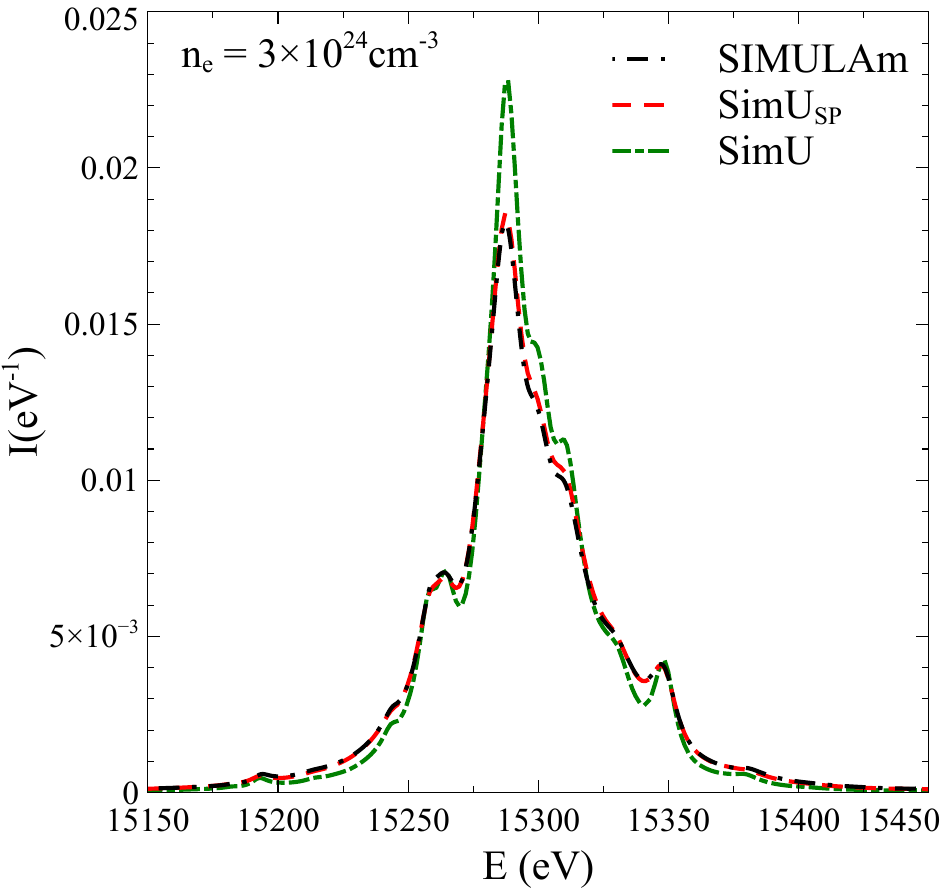}
}
\subfloat[\label{fig:HeBe2_1e25}]{
    \includegraphics[height=0.3\linewidth]{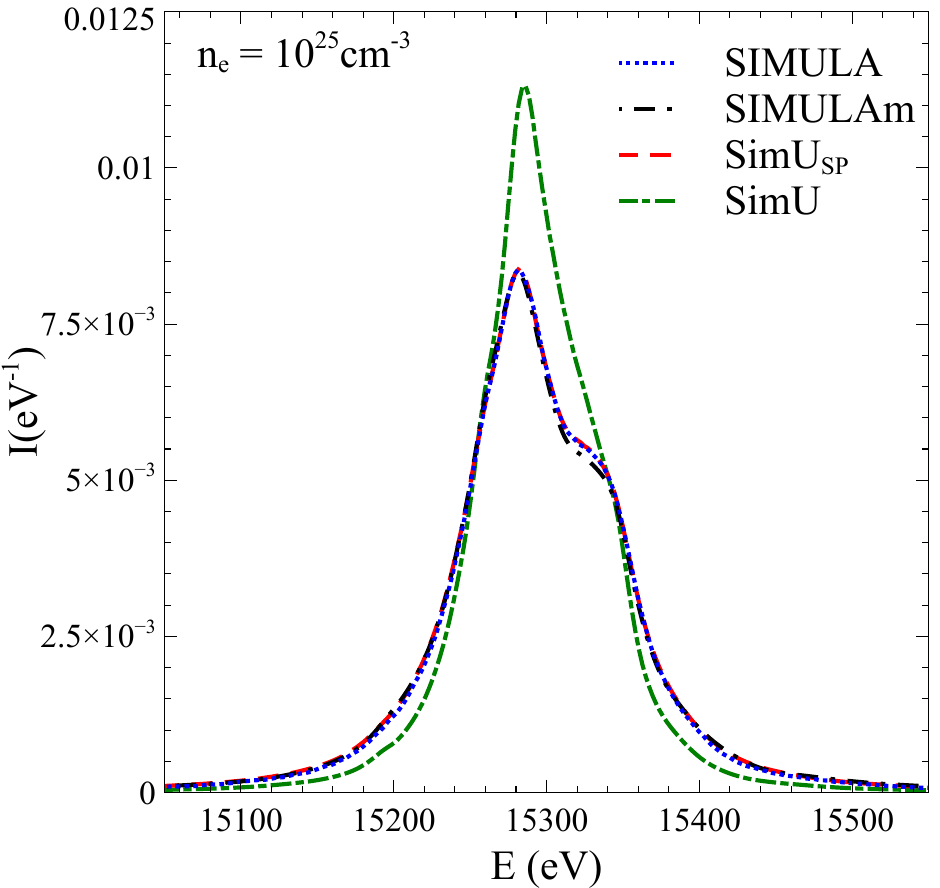}
} 
\hfill
\subfloat[\label{fig:HeBe3_1e24}]{
    \includegraphics[height=0.3\linewidth]{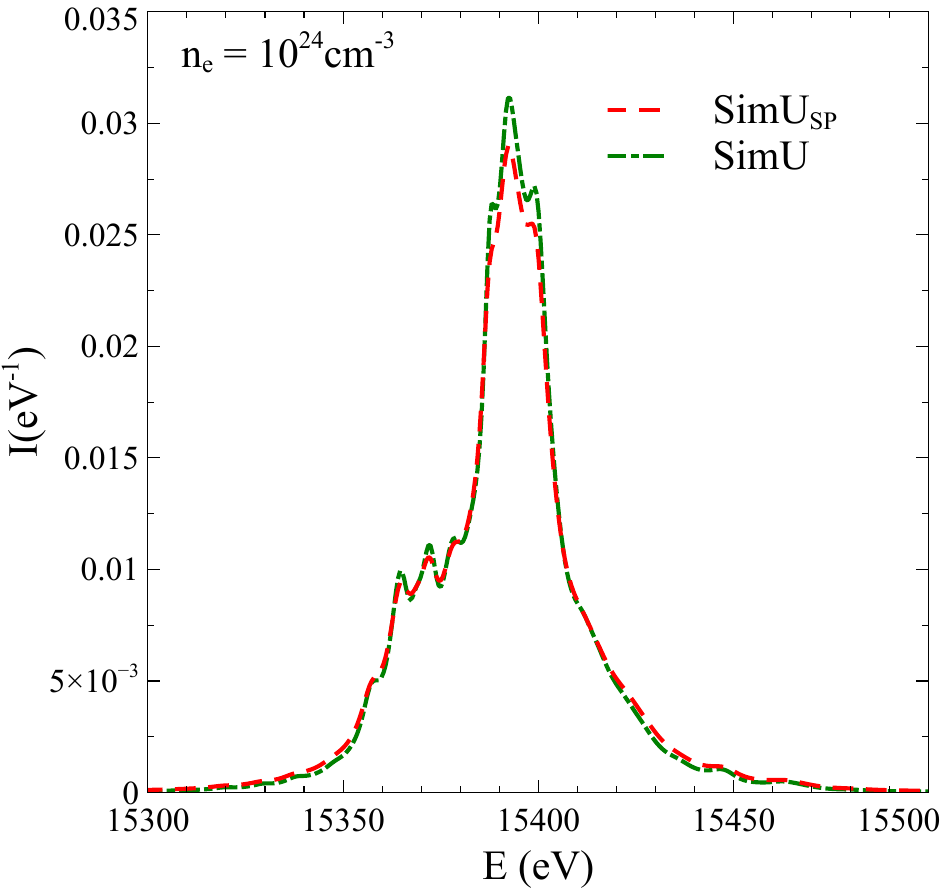}
} 
\subfloat[\label{fig:HeBe3_3e24}]{
    \includegraphics[height=0.3\linewidth]{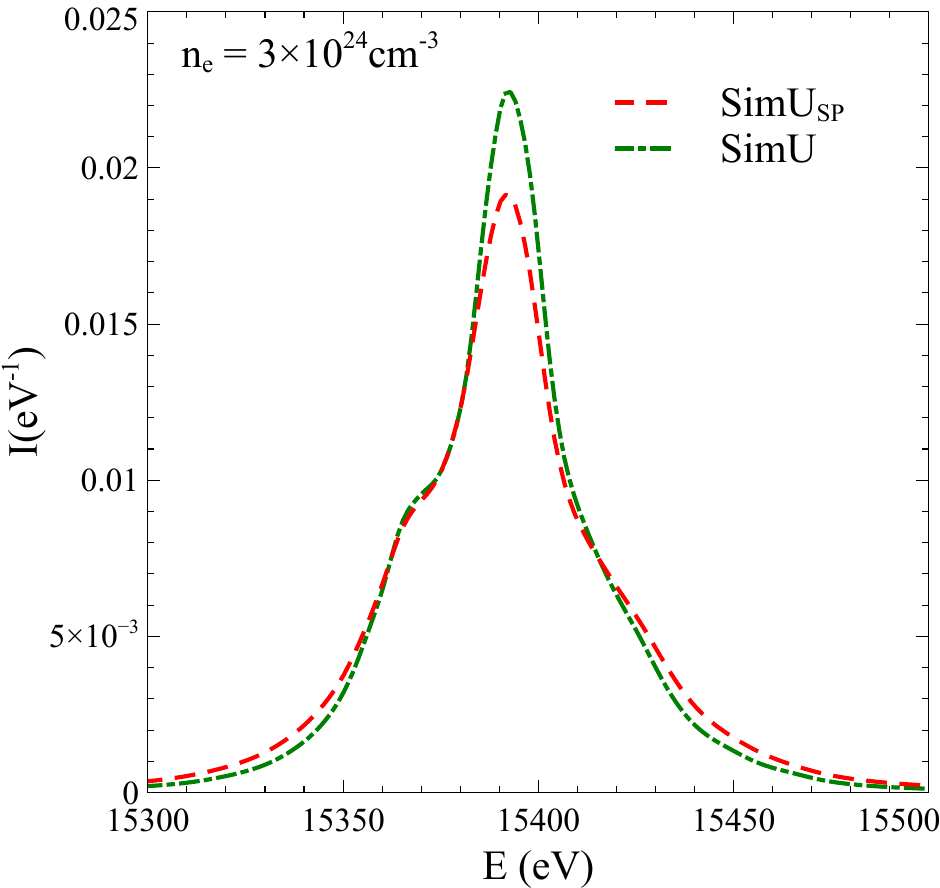}
}
\subfloat[\label{fig:HeBe3_1e25}]{
    \includegraphics[height=0.3\linewidth]{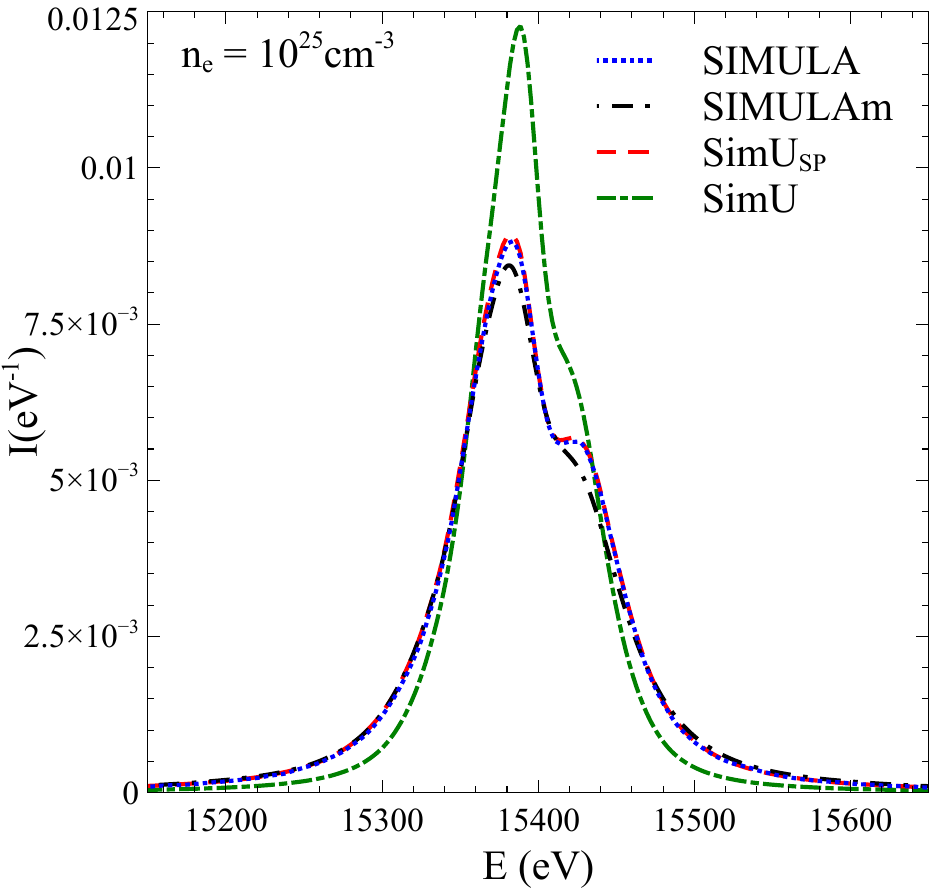}
} 
\caption{Same as \cref{fig:He}, but for $n=2$ (top) and $n=3$ (bottom) Li-like satellites. Figures \ref{fig:HeBe2_1e24} and \ref{fig:HeBe3_1e24} (left) correspond to $n_e=\qty{1e24}{\ccm}$, \ref{fig:HeBe2_3e24} and \ref{fig:HeBe3_3e24} (center) to $n_e=\qty{3e24}{\ccm}$, and \ref{fig:HeBe2_1e25} and \ref{fig:HeBe3_1e25} (right) to $n_e=\qty{1e25}{\ccm}$. $T=\qty{3}{keV}$ is assumed in all cases.}
\label{fig:Li}
\end{figure*}

For the lower densities, the two main components of the \HeB complex can be clearly resolved, namely, $^3P_1 \rightarrow$ $^1S_0$ at $\sim \qty{15409}{eV}$ and $^1P_1\rightarrow$ $^1S_0$ at $\sim \qty{15435}{eV}$---where we are naming the states using the $^{(2S+1)}L_J$ notation. Additionally, even for $n_e=\qty{1e24}{\ccm}$, the $^1D_2 \rightarrow$ $^1S_0$ forbidden transition starts to appear at $E\sim \qty{15445}{eV}$, owing to the Stark mixing of the states. This effect becomes clearer at $n_e = \qty{3e24}{\ccm}$. However, for the highest density case ($n_e=\qty{1e25}{\ccm}$), the spectra look significantly different as the contribution from Stark-mixed states becomes more intense than that of the unperturbed states. The main peaks that appear in the spectra correspond to two dipole-forbidden transitions: the aforementioned $^1D_2 \rightarrow$ $^1S_0$ at $\sim \qty{15445}{eV}$ and the additional $^3P_2 \rightarrow$ $^1S_0$ line at $E\sim\qty{15430}{eV}$, with the \textit{original} lines contributing only in their wings.

For all plasma densities, all models show qualitatively similar results. Note that SIMULA, SIMULAm, and SimU$_\text{SP}$ show a remarkable agreement, which is indicative of the fact that, despite the differences between the codes (both numerical and physical), when the particles are considered independent, all codes reproduce the same shape of the spectral line. On the other hand, it is clear that when interaction with the radiator is included (SimU), the line becomes narrower for all considered densities. This is not unexpected since this interaction causes the radiator to repel perturbing ions, decreasing the ionic microfield (which is a significant broadening agent in these conditions).

In particular, the agreement of SIMULAm, a hybrid code, with \glspl{CSM} like SIMULA and SimU$_\text{SP}$, is remarkable, considering that the electron dynamics are not explicitly resolved in SIMULAm, but instead introduced following the standard theory. This approach, which, as mentioned above, speeds the simulations by a factor of $\sim 50$, has, to our knowledge, not been presented previously---while a similar approach was used by \citet{stamm:1979a}, the electron effect was introduced as a whole in that case, rather than at every timestep, thus yielding different results owing to the non-commutation of the electron and ion evolution operators.

As mentioned in \cref{sec:Codes}, for the particular case of $n_e = \qty{1e25}{\ccm}$, we also performed calculations with the code DinMol which accounts for the full interaction between all particles. Interestingly, the full molecular dynamics simulations follow the same trend for the lineshape that was observed with the simpler \glspl{CSM}. That is, when all particles are allowed to interact with each other, the line narrows even further since all interactions are smoothed. However, as explained in Refs.~\cite{gigosos:2021, gonzalezherrero:2025}, tracking the full particle interaction enables us to account for the sudden loss of correlation when an emitter captures a free electron, known as \textit{recombination broadening}. The interplay between these two effects (recombination broadening and full particle interaction) results in a lineshape that lies between the predictions of SIMULA and SimU. This is shown in \cref{fig:HeB_DinMol}, where we show the line profile obtained for SIMULA, SimU, and DinMol with (w) and without (w/o) recombination broadening.

The results for the \HeB line width are summarized in \cref{fig:FWHA}, where we show the \acrfull{FWHA} of the line as a function of density for all codes presented here (\gls{FWHA} is a metric that is less sensitive to ion dynamics than the full width at half maximum \cite{gigosos:2003}). This shows again that when straight paths are assumed for the perturbers, all codes agree regardless of their computational differences. This figure also includes a line indicating a $n_e^{2/3}$ dependence of the line widths on the density typical for the quasistatic Stark effect of hydrogen-like transitions~\cite{griem:1974a}. It can be seen that for all models, the line widths closely follow this trend.

\subsection{\texorpdfstring{$n=2$}{n = 2} and \texorpdfstring{$n=3$}{n = 3} Li-like satellites}

Calculations of the $n=2$ and $n=3$ Li-like satellites are shown in \cref{fig:Li} (top and bottom, respectively), in the same manner as \cref{fig:He}. As for the case of the \HeB line, the spectral range has been adjusted for each subfigure. Owing to computational limitations, not all cases are presented for all codes. In particular, SIMULA results are presented only for the highest-density cases (most computationally affordable). Nevertheless, it should be borne in mind that, as discussed in the previous section, the differences between codes are mainly due to the type of perturber trajectories assumed.

We want to emphasize the computational complexity of these calculations. The number of states involved in these transitions is of the order of hundreds. Therefore, for each simulation timestep, the codes must work with matrices of $\sim 300\times300$ elements to evaluate the evolution operator, using algorithms that scale as $O(N^3)$, where $N$ is the matrix size. This alone requires significant computer power. Combined with the small size of the timestep necessary to accurately model the time-dependent electric field in plasma at these temperatures and densities, and the need to average over several simulation runs, this results in several thousand hours of computational time for each pair of temperature and density. The calculations presented here are, to our knowledge, the first computer simulations of the lineshape of these transitions, a crucial step for properly studying and modeling full experimental spectra.

\subsection{Complete spectrum of the He-\texorpdfstring{$\beta$}{beta} complex}

From the obtained lineshapes of the different components of the \HeB complex, it is possible to obtain a synthetic spectrum of the whole emission. To do so, we chose the same conditions used in Ref.~\cite{gallardo-diaz:2024a}, that is, a 1:1 mixture of $^3$He and deuterium with krypton as a minority at $T=\qty{3}{keV}$ and $n_e=\qty{1e24}{\ccm}$. 

We used the collisional-radiative code ABAKO~\cite{florido:2009} to solve the atomic kinetics for the temperature and density conditions of interest. Upon doing so, the normalized line profiles were used with ABAKO such that the emissivity and opacity of the corresponding transitions could be properly distributed across photon energies. Additionally, we included the thermal Doppler broadening (of \qty{4.3}{eV}) in the spectral calculations.

\begin{figure}
\centering
\includegraphics[width=\columnwidth]{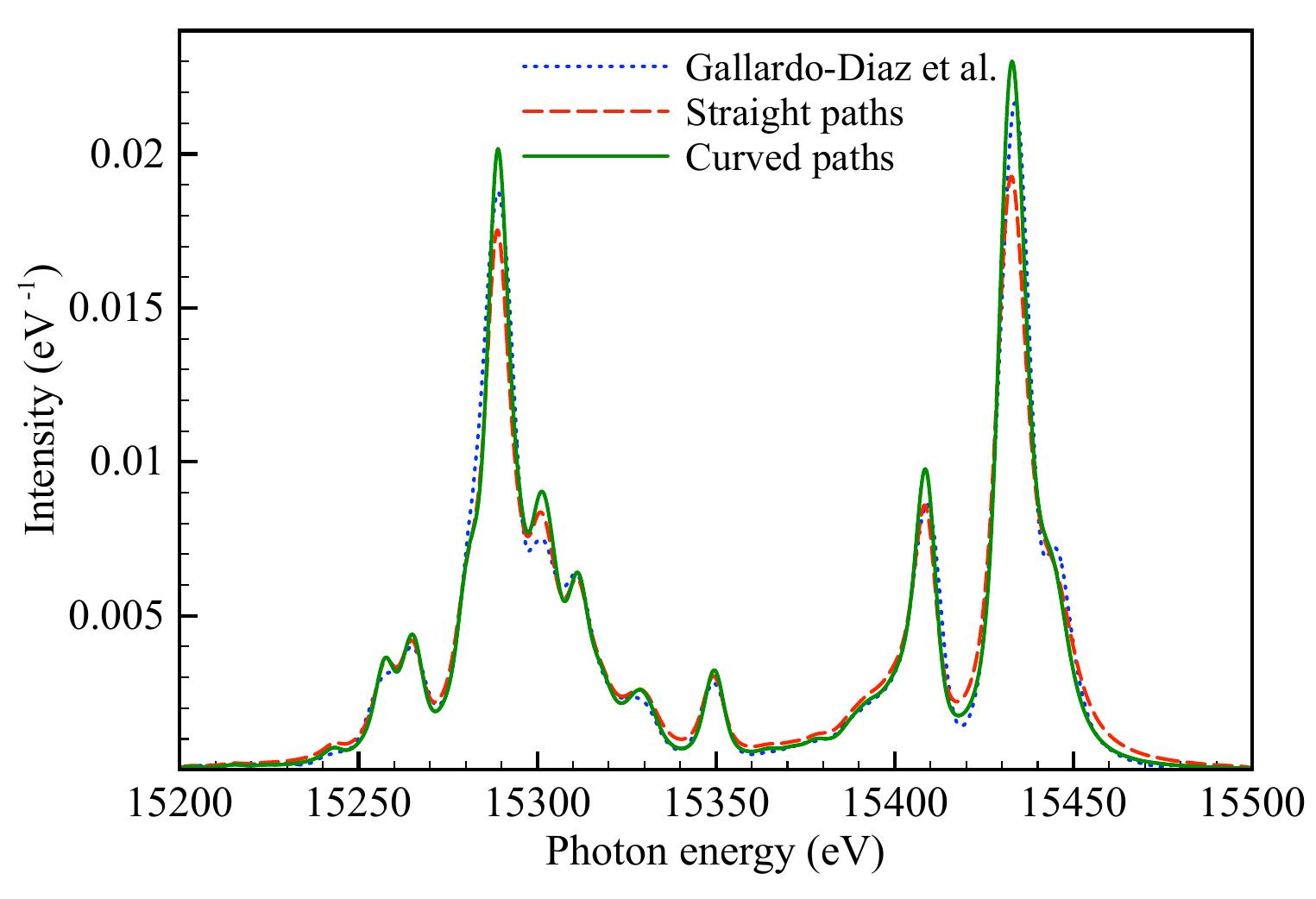}
\caption{Comparison of the full spectrum of the Kr \HeB complex given in a previous study~\cite{gallardo-diaz:2024a} with our results.
A $T=\qty{3}{keV}$, $n_e=\qty{1e24}{\ccm}$ plasma with a trace minority of Kr in a 1:1 atomic mixture of D and $^3$He is assumed. The spectra are area-normalized.}
\label{fig:spectra}
\end{figure}

The results of the full \HeB spectrum calculated with SimU and SimU$_{\text{SP}}$ (i.e., using curved and straight path trajectories respectively) are shown in \cref{fig:spectra} for the conditions above. We also show, for comparison, the resulting spectra using the line profiles from Ref.~\cite{gallardo-diaz:2024a}, where the spectral line shapes were modeled with the code MERL \cite{boercker1987, woltz1988, mancini1991} and the PrismSPECT model \cite{mcfarlane2013} was used to solve the atomic kinetics.
In order to remove any differences in the spectra arising from the different atomic kinetics, we used ABAKO in all cases with the lineshapes extracted from Fig.~11 of Ref.~\cite{gallardo-diaz:2024a}.

It can be seen that the MERL predictions generally lie between those for the simulations with curved and straight paths. However, MERL does not account for the broadening effect of ion dynamics, which is taken into consideration in the simulations. This effect is non-negligible at the conditions of interest and is likely the cause for the prediction of a \textit{shoulder} in the high-energy wing of the \HeB line (at $\sim \qty{15.45}{keV}$), which is not as prominent in the simulations.

\section{Simulation-Equivalent Interference Terms and their effect on the satellite emission}
\label{sec:IT}

\begin{figure*}
\centering
\subfloat[\label{fig:Interference_1e24}]{
    \includegraphics[width=0.45\linewidth]{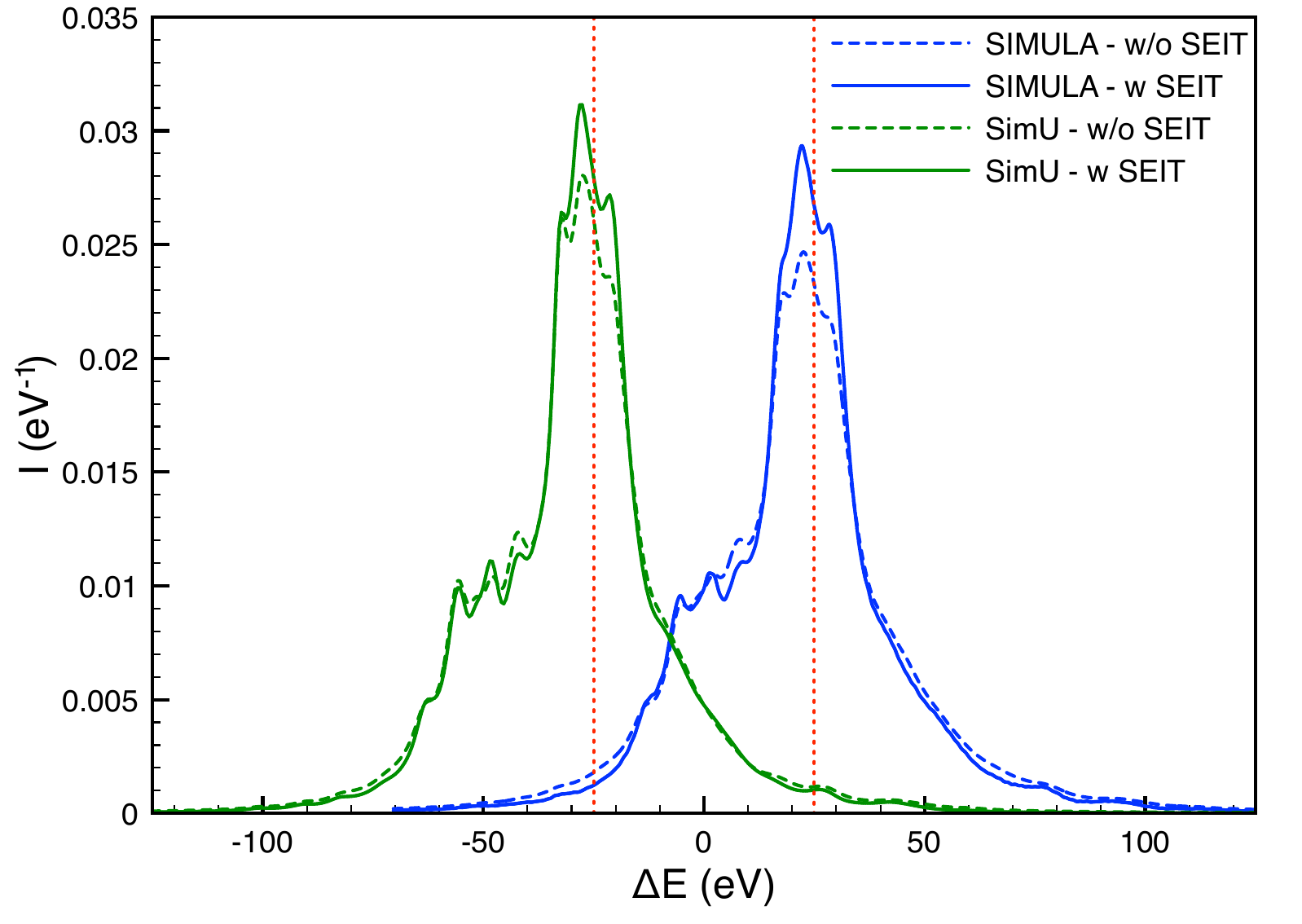}
}
\subfloat[\label{fig:Interference_1e25}]{
    \includegraphics[width=0.45\linewidth]{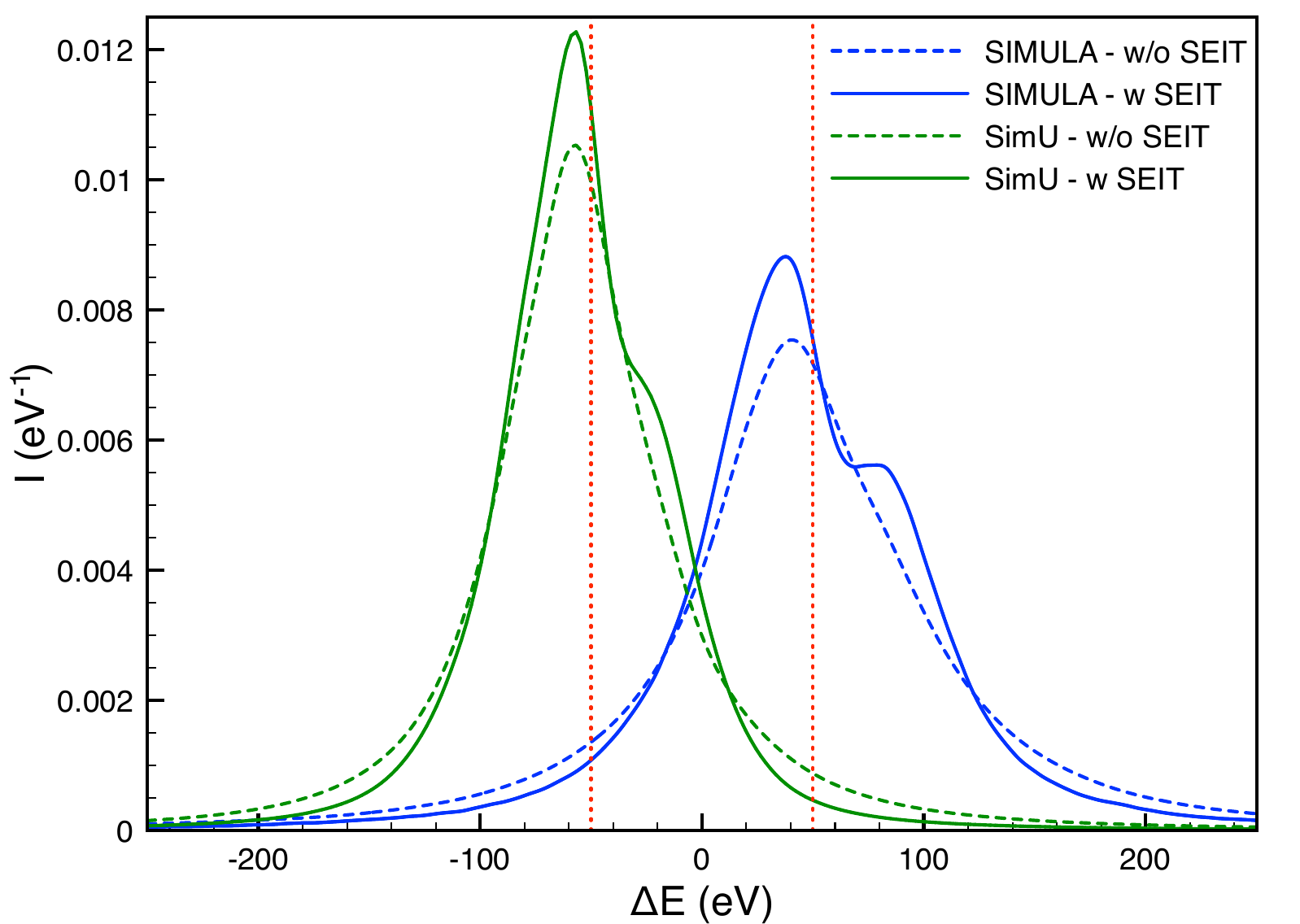}
}
\caption{Area-normalized line profiles of the $n=3$ Li-like satellite emission for electron densities of $\qty{1e24}{\ccm}$ (a) and $\qty{1e25}{\ccm}$ (b) with (w, solid lines) and without (w/o, dashed lines) including SEIT. The blue lines correspond to SIMULA, while the red lines correspond to SimU. It can be seen that the SEIT always broaden the line (following the same trend as the classic interference terms)}; however, this effect is more pronounced for the highest density considered. The lines for SIMULA and SimU have been respectively upshifted and downshifted to ease the view, with the center of the lines marked with red dotted lines.
\label{fig:InterferenceTerms}
\end{figure*}

With the results presented in the previous sections regarding the Li-like satellites, we are now in a position to discuss the effect of the so-called \textit{interference terms} on the lineshape of these complex transitions.
The name originates from a discussion~\cite{griem:1973, hey:1975, voslamber:1976, griem:1976} in the context of the electron impact broadening theory.
In this framework, the impact operator $\Phi$, determining the width and shift of spectral lines, contains off-diagonal elements with the $\vec{D}_a\cdot\vec{D}_b$ product, where $\vec{D}_a$ and $\vec{D}_b$ are the dipole moment operators of the upper and lower levels of a transition, respectively, and arise from the interaction of the atom with the plasma microfields.
These elements were called interference terms, as they represent a contribution to the spectral line shape caused by the correlation of the wavefunctions of the upper and lower levels evolving at separate rates.

Mathematically, the impact operator $\Phi$ can be written as a sum of three terms \cite{mancini1991, galtier:2013}, as

\begin{multline}
\Phi_{\alpha\beta ij}= \delta_{ij}\sum_{\gamma}\vec{\mathbf{d}}_{\alpha \gamma} \cdot \vec{\mathbf{d}}_{\gamma \beta}G(\Delta{\omega}_{\gamma i})\\
+\delta_{\alpha \beta}\sum_{k}\vec{\mathbf{d}}_{ik} \cdot \vec{\mathbf{d}}_{kj} \,
G(-\Delta{\omega}_{\alpha k})\\
- \vec{\mathbf{d}}_{\alpha \beta} \cdot \vec{\mathbf{d}}_{ji} \left[ G(\Delta{\omega}_{\alpha j}) + G(-\Delta{\omega}_{\beta i}) \right],
\label{eqn:impactTheory}
\end{multline}
where the Latin and Greek indices represent states from the upper and lower level of the transition, respectively. The function $G$ is related to the evolution operator of the emitter and the elements $\Delta\omega_{\alpha i} = \omega-\omega_{\alpha i}$, with $\omega_{\alpha i}$ corresponding to the transition frequency between the $\alpha$ and $i$ states.

Each of the terms in the sum corresponds to different transitions between the upper and lower levels. The first term involves transitions in which the dipole interaction causes the upper state to evolve by means of an intermediate state ($\alpha\rightarrow\gamma\rightarrow\beta$), but the lower state remains unchanged. Equivalently, the second term involves the evolution of the lower state ($i\rightarrow k\rightarrow j$) but not the upper state. Finally, the last term, involves a transition in which the upper and lower states both evolve separately ($\alpha\rightarrow\beta$ and $i\rightarrow j$). It is this third term that corresponds to the so-called \textit{interference terms}. Importantly, if $\alpha=\beta$ and $i=j$, the interference terms are zero (since $\vec{\mathbf{d}}$ is an odd operator).

In all computer simulation models of line-shape broadening, the effect of the interference terms is always included.
However, to our knowledge, no accepted approach exists~\cite{iglesias:2026a, gomez:2026a} on how to model the {\em omission} of this effect in a \gls{CSM}.
In the absence of any prior published study on this subject, we introduce the notion of \gls{SEIT} that retains the physical meaning of the interference terms in the impact approximation and is believed to produce qualitatively and, likely, quantitatively similar results.
Namely, from \cref{eqn:TimeDomain}, it can be seen that the shape of the spectral line is given by the Fourier transform of the trace of a matrix.
If the density matrix is equal to the unity matrix, this trace can be explicitly written as
\begin{multline}
    \label{eqn:IT}
    \text{Tr}\left[D(t)D(0)\right] = \sum_{ij\alpha\beta} U^\dagger_{ij}(t)D_{j\alpha}(0)U_{\alpha\beta}(t)D_{\beta i}(0) = \\
    =\sum_{ij\alpha\beta} U^\dagger_{ij}D_{j\alpha}U_{\alpha\beta}D_{\beta i}
\end{multline}
(the explicit time dependence was dropped for brevity).
Physically, this involves the transitions $j\rightarrow\alpha$ and $\beta\rightarrow i$, coupled by the evolution operator mixing the states $i,j$ and $\alpha,\beta$, respectively.

Following the philosophy from Eqn. \ref{eqn:impactTheory}, the \gls{SEIT} are defined as the terms in the sum that involve transitions with different upper and lower states, that is
\begin{equation}
    \text{SEIT} = \sum_{ij\alpha\beta} (1-\delta_{ij})(1-\delta_{\alpha\beta})U^\dagger_{ij}D_{j\alpha}U_{\alpha\beta}D_{\beta i}.
\end{equation}
In practice, it is more convenient to include in the definition of \gls{SEIT} the terms corresponding to transitions in which either the upper state or the lower state has evolved (not necessarily both of them). While these terms are not, strictly speaking, \textit{of interference}, they are negligible \cite{galtier:2013}, and so, in practice, including them in the definition of \gls{SEIT} makes no difference in the obtained spectra (but simplifies the computation). Therefore, for practical purposes, we define the \gls{SEIT} as
\begin{equation}
    \text{SEIT} = \sum_{ij\alpha\beta} (1-\delta_{ij}\delta_{\alpha\beta})U^\dagger_{ij}D_{j\alpha}U_{\alpha\beta}D_{\beta i}.
\end{equation}
In doing so, we can remove the effect of \gls{SEIT} by only considering transitions going from the same upper state to the same lower state, i.e., those elements with $i=j$ and $\alpha=\beta$.
This means that \cref{eqn:IT} reduces to the ``SEIT-free trace'' form as
\begin{equation}
    \text{Tr}\left[D(t)D(0)\right]_{\text{no SEIT}} = \sum_{i\alpha} U^\dagger_{ii}D_{i\alpha}U_{\alpha\alpha}D_{\alpha i},
\end{equation}
involving only the diagonal elements of the evolution operator $U$. \footnote{It is important to note that, using this definition, the trace of the dipole correlation function removes intermediate states in the transition from a state in the upper level to one in the lower level. This effectively eliminates the \textit{quantum interference} caused by the summation of different amplitudes of the evolution operator.}
It is stressed that the evolution operator $U(t)$ is correctly and fully calculated at every timestep of the simulation;
only some terms from the trace summation in \cref{eqn:TimeDomain} [or equivalent ones in \cref{eqn:FreqDomain}] are omitted.

\begin{figure}
\includegraphics[width=\columnwidth]{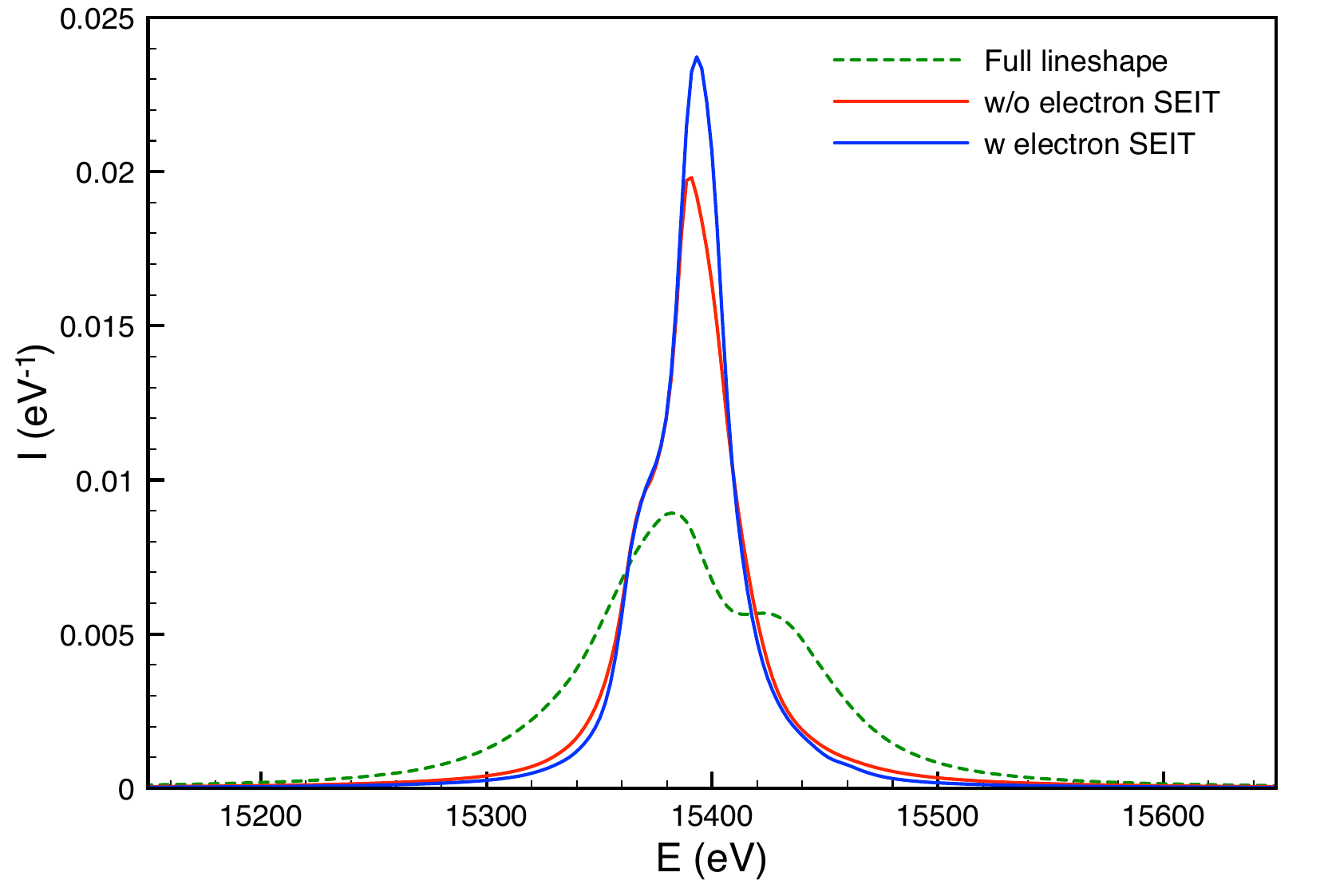}
\caption{Area-normalized line profiles from SimU$_{\text{SP}}$ of the $n=3$ Li-like satellite emission for an electron density of $\qty{1e25}{\ccm}$, considering only electron broadening with (w) and without (w/o) \gls{SEIT}. The full lineshape (including both ions and electrons) shown in Fig. \ref{fig:HeBe3_1e25} has been added to better illustrate this effect.}
\label{fig:OnlyElectrons}
\end{figure}

Using this definition, we studied the effects of the SEIT on the shape of the Li-like satellites.
It is found that their effect is almost negligible for the $n=2$ satellites, but it is not the case for the $n=3$ satellites.
For these, at the lowest density considered ($n_e=\qty{1e24}{\ccm}$), the effect of the \gls{SEIT} is observable, although negligible for practical applications, whereas for the highest density ($n_e=\qty{1e25}{\ccm}$) the \gls{SEIT} effect significantly modifies the shape of the spectral line.
In this case, removing the \gls{SEIT} caused the satellite emission to lose its structure, such as the shoulder on the high-energy wing of the line.
In all cases, removing the \gls{SEIT} resulted in a broadening of the spectral emission, in agreement with previous findings within the framework of electron impact theory~\cite{iglesias:2010a}.
This can be seen in \cref{fig:InterferenceTerms}, where we show the result of removing the \gls{SEIT} for the highest and lowest densities of consideration using both SimU and SIMULA (curved and linear particle trajectories, respectively).

Importantly, since in the full computer simulations the effects of ions and electrons are considered jointly, this approach is not equivalent to that of removing the interference terms in the electron impact broadening theory
within the ``standard theory'' approach, where ions are always considered in the quasistatic approximation.
For a more direct comparison, we ran computer simulations only with electrons.
An example of the results obtained in these calculations is shown in \cref{fig:OnlyElectrons} for the $n=3$ satellites and electron density of $\qty{1e25}{\ccm}$ from SimU$_{\text{SP}}$, where the effect of removing the \gls{SEIT} is clearly observed.

\begin{figure}
\includegraphics[width=\columnwidth]{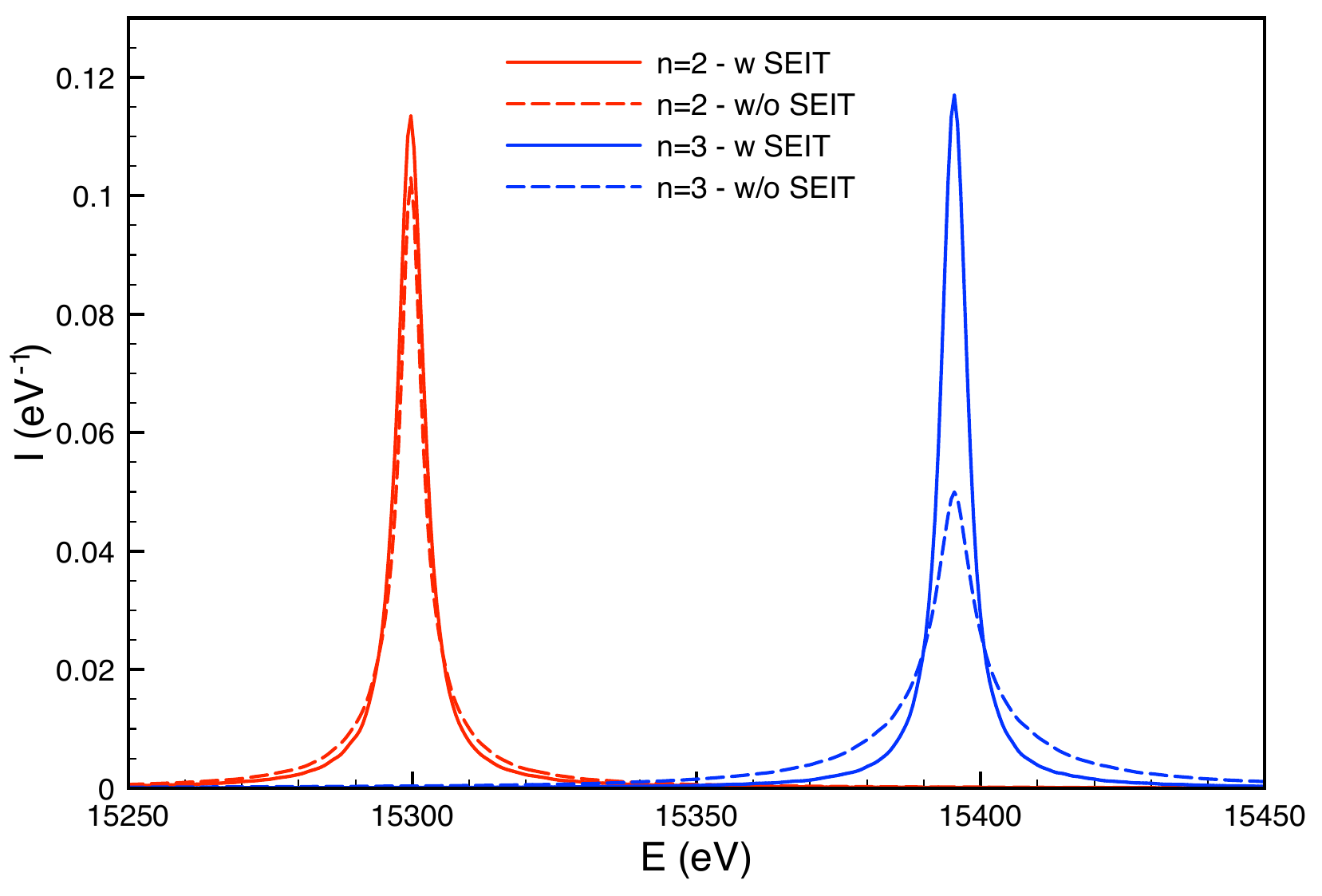}
\caption{The effect of removing the \gls{SEIT} in the electron broadening assuming fully degenerate upper and lower level energies of the $n=2$ and $n=3$ Li-like satellite transitions, for an electron density of \qty{1e25}{\ccm}.}
\label{fig:Degenerate}
\end{figure}

The dashed green line in Fig. \ref{fig:OnlyElectrons} shows the full line profile (as shown in Fig. \ref{fig:HeBe3_1e25}) for comparison.
Note that for these conditions, the electron broadening is minor compared to that of ions. 
Furthermore, even that minor contribution is obscured by the energy structure of the numerous atomic levels, indicated by the asymmetric lineshapes in \cref{fig:OnlyElectrons}.
To further isolate the effect of the electron broadening and the importance of the \gls{SEIT} on it, the calculations were repeated assuming fully degenerate upper and lower levels of the satellite transitions.
The results are shown in \cref{fig:Degenerate}.
There, it is clearly seen that the \gls{SEIT} are very important for the electron broadening of the $n=3$ satellites, affecting the linewidth by a factor of two. Qualitatively similar results are obtained for lower densities.

This analysis explains the seemingly contradicting results of \citet{gallardo-diaz:2024a} who concluded that the interference terms played a minor role in the calculation of the $n=3$ Li-like satellite emission of the Kr \HeB line at $n_e=\qty{1e24}{\ccm}$.
Indeed, the effect appears to be small, but only because the electron impact broadening, whether with or without the interference terms, is minor compared to the other line-shape formation factors (ion broadening and atomic structure).
However, under different conditions, such as significantly higher densities or transitions from levels with higher principal quantum numbers, this may no longer be the case, and omitting the interference terms would result in significant errors.

\section{Conclusions and future work}
\label{sec:Conclusion}

We have presented computer simulations of the Stark-broadened shape of the krypton \HeB line, as well as, for the first time, the associated $n=2$ and $n=3$ Li-like satellites at \gls{ICF} conditions ($T=\qty{3}{keV}$, $n_e=\qtyrange{1e24}{1e25}{\ccm}$). These results demonstrate the robustness of current simulation codes and the utility of this technique for \gls{HED} spectroscopic diagnostics. All codes agree qualitatively and show that the results only depend on the simulation's complexity level. The differences between different codes caused by the treatment of particle trajectories become more pronounced at higher densities, as the interaction between the emitter and perturbers in the plasma plays a more significant role in the field dynamics.

We have presented SIMULAm, a hybrid code combining computer simulations and standard theory, which shows remarkable agreement with the full simulation codes SIMULA and SimU$_\text{SP}$, speeding the simulation up by a factor of $\sim 50$. Future work includes testing the approach used by SIMULAm with curved trajectories.

Additionally, we have defined the \textit{simulation-equivalent interference terms } (SEIT) based on the concept of \textit{interference terms} from the electron impact broadening theory and analyzed their effect on the shapes of the Li-like satellites. While the effect is practically negligible for the $ n=2$ satellites, the $n=3$ satellites are noticeably affected. Therefore, omitting these terms in calculations based on the electron impact theory (for the sake of a significant speed-up) may result in inaccuracies that tend to increase with higher densities.

Future work on this topic should extend the current results beyond the dipole approximation by accounting for all-order full plasma--radiator interaction and studying the effects of penetrating electron and ion collisions \cite{stambulchik:2022a, gomez:2024b}.
In particular, these effects should result in the plasma polarization shift of the Kr \HeB complex that appears to be observed at \gls{NIF}~\cite{hill:2022a}.

\section*{Data availability statement}

The data that support the findings of this article are openly available \cite{data_Zenodo}. 

\begin{acknowledgments}

The authors would like to acknowledge discussions with C.~A.~Iglesias and T.~A.~Gomez on the topic of interpreting effects of interference terms in computer simulations.

M.A.G., G.P.-C., and R.F. acknowledge the support of Research Grant No. PID2022-137632OB-I00 from the Spanish Ministry of Science and Innovation.

The work of M.A.G., G.P.-C., and R.F. has also been carried out within the framework of the EUROfusion consortium, funded by the European Union via the Euratom Research and Training Program (Grant Agreement Nos. 633053 and 101052200—EUROfusion)
The views and opinions expressed are, however, those of the author(s) only and do not necessarily reflect those of the European Union or the European Commission. Neither the European Union nor the European Commission can be held responsible for them. The involved teams have operated within the framework of the Enabling Research Projects: Grant numbers AWP21-ENR-IFE.01.CEA and AWP24-ENR-IFE.02.CEA-01.

The work of E.S. was partly supported by the Lawrence Livermore National Laboratory (USA) and by the University of Michigan Multi-University Center of Excellence for Magnetic Acceleration, Compression, and Heating (MACH).

This study was inspired by a computational challenge considered at the 6th Spectral Line Shapes in Plasmas Code Comparison Workshop~\cite{slsp6}. We acknowledge the support of the International Atomic Energy Agency (IAEA) in organizing the meeting.

\end{acknowledgments}

\bibliography{biblio}

\end{document}